# Electronic Transport through QD in the whole temperature range including both the high- and the low-T limits with the equation-of-motion technique

Kuk-Chol Ri, Chol-Won Ri, Gum-Hyok Jong

*Department of Physics, Kim Il Sung University, Pyongyang, DPR Korea*

We have studied theoretically the Kondo effect in the quantum dot(QD) within the whole range of temperature by using the equation-of-motion(EOM) technique based on the non-equilibrium Green function formalism. We have taken the finiteness of Coulomb correlation and the non-equilibrium effect into account by calculating the correlation terms emerged from the decoupling approximation using EOM method for the lesser Green function. We showed that the result is in good qualitative agreement with the results of NCA, NRG and NRPT, etc. , even using EOM method which is being recognized as a 'conventional' method . The results are the generalization into the pseudo-equilibrium state of the Refs. 32,33 and can be used to describe a non-equilibrium state under the bias voltage which is not so large.

## I. Introduction

The Kondo effect[1,2] in electronic transport through QD strongly coupled to metallic leads was predicted theoretically[3,4] and also observed experimentally[5-11]. The Kondo effect is a typical instance to demonstrate an importance of the many-body effect.

The main features of the Kondo effect in QD are the appearances of the Kondo peak in the density of state (DOS) at the Fermi level and the zero-bias maximum of the differential conductance. Exchange coupling of the electron located on the dot with conduction electrons of the leads gives rise to screening of the dot spin, which consequently leads to the Kondo resonance at sufficiently low temperatures.

Several theoretical techniques have been developed to study this intriguing phenomenon, including a Green function technique based on the EOM or diagrammatic methods, slave-boson mean field theory, scaling approach, numerical renormalization group techniques(NRG) and real time renormalization group techniques(RTRG), etc.[12-28]

One of the techniques used in the non-equilibrium situations is the non-equilibrium Green function(NGF) techniques.[15,21,22,25,26] To calculate DOS and electric current one needs both retarded and lesser Green functions. The former one is usually calculated by the EOM method in the framework of some decoupling approximation schemes. The most commonly used approximation is the one worked out by Meir *et al*[15]. This approximation describes the situation rather well for temperatures close to the Kondo temperature $T_K$ as well as above $T_K$. However, it is not correct for much smaller temperatures $T \ll T_K$. In this regime, the approximation developed by Lacroix[29] is more appropriate.

Most of literatures(e.g., Refs. 21,25,26,28,31,etc) based on the EOM method used the decoupling approximation of Meir *et al*[15], and so they haven't considered the very low temperature region $T \ll T_K$. On the other hand, the Kondo effect in the whole range of temperature including $T_K$ was studied in Refs. 32,33 by using the approach developed by Lacroix[29]. However, they[32,33] calculated the correlation functions using the fluctuation-dissipation theorem, which is valid only for the equilibrium state.

In the case of non-equilibrium problem, the EOM method is not so commonly used because of some difficulties with calculating the lesser Green function was derived on the basis of some heuristic assumptions. Most of references (e.g., Refs. 21,25,26,28,31, etc) have adopted the Ng *ansatz*[30] to calculate the self-energy, whereas in Ref. 31 both the lesser and retarded Green functions were calculated consistently within the same approximation scheme not using the Ng *ansatz*. However, such approximations for the lesser Green function do not conserve charge current in asymmetrical systems.

In this paper, we are going to investigate theoretically the Kondo effect in QD's with finite Coulomb correlation U in the whole range of temperature including the Kondo temperature $T_K$ by using the EOM method based on NGF technique. Our goal is to show how to obtain the DOS and the conductance which agree well with the results of NCA,NRG and NRPT under the bias voltage that is not so large, by using the EOM technique.

The paper is organized as follows. In Sec. II we describe the model and establish the theoretical formalism so that DOS and the electric current in the non-equilibrium state cam be calculated in the whole range of temperature by applying the EOM method and the decoupling approximations to the Green function. The numerical results for DOS and the differential conductance are presented and discussed in Sec. III. Finally, the summary and the general conclusions are in Sec. IV.

## II. Theoretical Formalism

We consider a single-level QD coupled to metallic leads by tunneling barriers. The whole system can be described by the following model Hamiltonian[31]:

$$H = H_{res} + H_D + H_T \tag{1}$$

The term $H_{res}$ in Eq. (1) describes the left ($\beta = L$) and right ($\beta = R$) leads:

$$H_{res} = \sum_{k\beta\sigma} \varepsilon_{k\beta\sigma} c^\dagger_{k\beta\sigma} c_{k\beta\sigma} \tag{2}$$

Where $\varepsilon_{k\beta\sigma}$ is the single-electron energy in the $\beta = L/R$ lead for the wave vector $k$ and electron spin $\sigma$, whereas $c^\dagger_{k\beta\sigma}$ and $c_{k\beta\sigma}$ denote the corresponding creation and annihilation operators.

The term $H_D$ in Eq. (1) describes QD and takes the form of

$$H_D = \sum_\sigma \varepsilon_\sigma d^\dagger_\sigma d_\sigma + U d^\dagger_\uparrow d_\uparrow d^\dagger_\downarrow d_\downarrow \tag{3}$$

,where $\varepsilon_\sigma$ denotes the energy of the discrete dot level of an electron with spin $\sigma$, $d^+_\sigma$ and $d_\sigma$ are the corresponding creation and annihilation operators, and $U$ denotes Coulomb correlation parameter.

The last term $H_T$ in Eq. (1) describes the tunneling between QD and leads, and is of the form

$$H_T = \sum_{k\beta\sigma} T_{k\beta\sigma} c^{\dagger}_{k\beta\sigma} d_\sigma + H.C. \tag{4}$$

, where $T_{k\beta\sigma}$ are the components of the tunneling matrix and describe the coupling between QD and $\beta$-lead.

The general expression for the electric current $I$ flowing from the left to the right electrode was derived by Jauho et al[25,26], who showed that in a stationary state the current $I$ is given by the formula.

$$I = \frac{ie}{2\hbar} \int \frac{dE}{2\pi} Tr\{[\mathbf{\Gamma}^L(E) - \mathbf{\Gamma}^R(E)]\mathbf{G}^<(E) + [f_L(E)\mathbf{\Gamma}^L(E) - f_R(E)\mathbf{\Gamma}^R(E)][\mathbf{G}^r(E) - \mathbf{G}^a(E)]\} \tag{5}$$

where $\mathbf{G}^\eta(E)$ ($\eta = r, a, <$) are the Fourier transforms $G^\eta_{\sigma\sigma}(E) \equiv \langle\langle d_\sigma; d^{\dagger}_\sigma \rangle\rangle^\eta_E$ of the non-equilibrium retarded, advanced and lesser Green functions of QD, $f_\beta(E)$ is the Fermi distribution function in the $\beta$-lead, and $\mathbf{\Gamma}^\beta(E)$ describes a contribution to the half-width of the dot level due to width of the dot level due to the tunneling through the $\beta$-barrier. $\mathbf{G}^\eta(E)$ ($\eta = r, a, <$) and $\mathbf{\Gamma}^\beta(E)$ are $2\times 2$ matrices in spin space, but if leads are nonmagnetic and we don't take the spin-flip process into account, $\mathbf{\Gamma}^\beta(E)$ would be diagonalized and so take the form of

$$\Gamma^\beta_{\sigma\sigma}(E) \equiv \Gamma^\beta_\sigma(E) = 2\pi \sum_k T^*_{k\beta\sigma} T_{k\beta\sigma} \delta(E - \varepsilon_{k\beta\sigma}) \tag{6}$$

Generally, $\Gamma^\beta_\sigma(E)$ are functions with the energy dependence, but we can suppose that they are constant in the conduction band[23,24,26]. Then, if $\Delta^\sigma_{min}$ and $\Delta^\sigma_{max}$ are the minimum edge energy and the maximum edge energy of the conduction band for spin $\sigma$ respectively, we take the following relation

$$\Gamma^\beta_\sigma(E) = \begin{cases} \Gamma^\beta_\sigma & (\Delta^\sigma_{min} \leq E \leq \Delta^\sigma_{max}) \\ 0 & (E < \Delta^\sigma_{min}, \Delta^\sigma_{max} < E) \end{cases} \tag{7}$$

In order to calculate the current Eq. (5), Green functions $G^{r(a)}_{\sigma\sigma}(E)$ and $G^<_{\sigma\sigma}(E)$ must be calculated.

## III. EOM Method

The equation of motion for the dot causal Green function $G_{\sigma\sigma}(E) = \langle\langle d_\sigma; d^{+}_\sigma \rangle\rangle_E$ is as follows.

$$(E - \varepsilon_\sigma)\langle\langle d_\sigma; d^{+}_\sigma \rangle\rangle_E = 1 + U\langle\langle d_\sigma n_{\bar{\sigma}}; d^{+}_\sigma \rangle\rangle_E + \sum_{k\beta} T^*_{k\beta\sigma} \langle\langle c_{k\beta\sigma}; d^{+}_\sigma \rangle\rangle_E \tag{8}$$

$$(E - \varepsilon_\sigma) G_{\sigma\sigma}(E) = 1 + U G^{(2)}_{\sigma\sigma}(E) + \sum_{k\beta} T^*_{k\beta\sigma} K_{k\beta\sigma,d\sigma}(E)$$

, where $G^{(2)}_{\sigma\sigma}(E) \equiv \langle\langle d_\sigma n_{\bar{\sigma}}; d^{+}_\sigma \rangle\rangle_E$ is the QD second Green function and $K_{k\beta\sigma,d\sigma}(E) \equiv \langle\langle c_{k\beta\sigma}; d_\sigma^+ \rangle\rangle_E$ is the electrode-QD Green function. The retarded (advanced) Green function $G_{\sigma\sigma}^{r(a)}(E)$ can be obtained as $G_{\sigma\sigma}^{r(a)}(E) = G_{\sigma\sigma}(E \pm i0^+)$ from the above causal Green function $G_{\sigma\sigma}(E)$.

The equation of motion for the electrode-QD Green function $K_{k\beta\sigma,d\sigma}(E) = \langle\langle c_{k\beta\sigma}; d_\sigma^+ \rangle\rangle_E$ is given as following.

$$(E - \varepsilon_{k\beta\sigma})\langle\langle c_{k\beta\sigma}; d_\sigma^+ \rangle\rangle_E = T_{k\beta\sigma}\langle\langle d_\sigma; d_\sigma^+ \rangle\rangle_E \qquad (9)$$
$$K_{k\beta\sigma,d\sigma}(E) = g_{k\beta\sigma}(E)T_{k\beta\sigma}G_{\sigma\sigma}(E)$$

The analytic continuation of the Eq. (9) is used to give the equation of motion for the lesser Green function of the electrode-QD $K_{k\beta\sigma,d\sigma}(E) = \langle\langle c_{k\beta\sigma}; d_\sigma^+ \rangle\rangle_E$

$$K_{k\beta\sigma,d\sigma}^<(E) = T_{k\beta\sigma}[g_{k\beta\sigma}^r G_{\sigma\sigma}^< + g_{k\beta\sigma}^< G_{\sigma\sigma}^a] \qquad (10)$$

, where $G_{\sigma\sigma}^{(2)r}(E) \equiv \langle\langle d_\sigma n_{\bar\sigma}; d_\sigma^\dagger \rangle\rangle_E^r$ is the second-order $g_{k\beta\sigma}^r(E) = (E - \varepsilon_{k\beta\sigma} + i0^+)^{-1}$ and $g_{k\beta\sigma}^<(E) = 2\pi i f_\beta(E)\delta(E - \varepsilon_{k\beta\sigma})$ denote the retarded and lesser Green function of the non-coupled electrode when there is no Coulomb interaction respectively.

Substituting Eq. (9) into Eq. (8) leads to the equation of motion for the QD retarded Green function $G_{\sigma\sigma}^r(E) = \langle\langle d_\sigma; d_\sigma^+ \rangle\rangle_E^r$

$$[E - \varepsilon_\sigma - \Sigma_{0\sigma}^r(E)]G_{\sigma\sigma}^r(E) = 1 + UG_{\sigma\sigma}^{(2)r}(E) \qquad (11)$$

, where $\Sigma_{0\sigma}^r$ is the self-energy in the absence of Coulomb interaction and describes a coupling to electrodes, and is defined as

$$\Sigma_{0\sigma}^r(E) \equiv \sum_{k\beta}|T_{k\beta\sigma}|^2 \, g_{k\beta\sigma}^r(E) = \sum_{k\beta}\frac{|T_{k\beta\sigma}|^2}{E - \varepsilon_{k\beta\sigma} + i0^+} \qquad (12)$$

On the other hand, the equation of motion for the second-order QD Green function $G_{\sigma\sigma}^{(2)}(E) = \langle\langle d_\sigma n_{\bar\sigma}; d_\sigma^+ \rangle\rangle_E$ can be written as

$$(E - \varepsilon_\sigma - U)\langle\langle d_\sigma n_{\bar\sigma}; d_\sigma^+ \rangle\rangle_E = \langle n_{\bar\sigma} \rangle$$
$$+ \sum_{k\beta}\left[T_{k\beta\sigma}^*\langle\langle c_{k\beta\sigma} n_{\bar\sigma}; d_\sigma^+ \rangle\rangle_E + T_{k\beta\bar\sigma}\langle\langle c_{k\beta\bar\sigma}^+ d_\sigma d_{\bar\sigma}; d_\sigma^+ \rangle\rangle_E - T_{k\beta\bar\sigma}^*\langle\langle c_{k\beta\bar\sigma} d_{\bar\sigma}^+ d_\sigma; d_\sigma^+ \rangle\rangle_E\right]$$

(13)

$$(E - \varepsilon_\sigma - U)G_{\sigma\sigma}^{(2)}(E) = \langle n_{\bar\sigma} \rangle + \sum_{k\beta}\left[T_{k\beta\sigma}^* K_{k\beta\sigma,d\sigma}^{A(2)}(E) + T_{k\beta\bar\sigma} K_{k\beta\sigma,d\sigma}^{B(2)}(E) - T_{k\beta\bar\sigma}^* K_{k\beta\sigma,d\sigma}^{C(2)}(E)\right]$$

Applying the EOM method to $G_{\sigma\sigma}^{(2)}$, second-order electrode-QD Green functions come out and the equation of motion for the second-order QD lesser Green function $G_{\sigma\sigma}^{(2)}$ contains these new second-order electrode-QD Green functions. These second-order Green functions are defined as follows.

$$K_{k\beta\sigma,d\sigma}^{A(2)}(E) \equiv \langle\langle c_{k\beta\sigma}n_{\bar{\sigma}};d_{\sigma}^{+}\rangle\rangle_{E}$$
$$K_{k\beta\sigma,d\sigma}^{B(2)}(E) \equiv \langle\langle c_{k\beta\bar{\sigma}}^{+}d_{\sigma}d_{\bar{\sigma}};d_{\sigma}^{+}\rangle\rangle_{E} \quad (14)$$
$$K_{k\beta\sigma,d\sigma}^{C(2)}(E) \equiv \langle\langle c_{k\beta\bar{\sigma}}d_{\bar{\sigma}}^{+}d_{\sigma};d_{\sigma}^{+}\rangle\rangle_{E}$$

Applying the EOM method to the second electrode-QD Green functions yields the following equations.

$$\begin{aligned}\left(E-\varepsilon_{k\beta\sigma}\right)\langle\langle c_{k\beta\sigma}n_{\bar{\sigma}};d_{\sigma}^{+}\rangle\rangle_{E} = \\ T_{k\beta\sigma}\langle\langle d_{\sigma}n_{\bar{\sigma}};d_{\sigma}^{+}\rangle\rangle_{E} + \sum_{q\alpha}\left[T_{q\alpha\bar{\sigma}}^{*}\langle\langle c_{k\beta\sigma}d_{\bar{\sigma}}^{+}c_{q\alpha\bar{\sigma}};d_{\sigma}^{+}\rangle\rangle_{E} - T_{q\alpha\bar{\sigma}}\langle\langle c_{k\beta\sigma}c_{q\alpha\bar{\sigma}}^{+}d_{\bar{\sigma}};d_{\sigma}^{+}\rangle\rangle_{E}\right]\end{aligned} \quad (15)$$

$$\begin{aligned}\left(E+\varepsilon_{k\beta\bar{\sigma}}-\varepsilon_{\sigma}-\varepsilon_{\bar{\sigma}}-U\right)\langle\langle c_{k\beta\bar{\sigma}}^{+}d_{\sigma}d_{\bar{\sigma}};d_{\sigma}^{+}\rangle\rangle_{E} = -\langle c_{k\beta\bar{\sigma}}^{+}d_{\bar{\sigma}}\rangle \\ T_{k\beta\bar{\sigma}}^{*}\langle\langle d_{\sigma}n_{\bar{\sigma}};d_{\sigma}^{+}\rangle\rangle_{E} + \sum_{q\alpha}\left[T_{q\alpha\sigma}^{*}\langle\langle c_{k\beta\bar{\sigma}}^{+}c_{k\beta\sigma}d_{\bar{\sigma}};d_{\sigma}^{+}\rangle\rangle_{E} + T_{q\alpha\bar{\sigma}}^{*}\langle\langle c_{k\beta\bar{\sigma}}^{+}d_{\sigma}c_{q\alpha\bar{\sigma}};d_{\sigma}^{+}\rangle\rangle_{E}\right]\end{aligned} \quad (16)$$

$$\begin{aligned}\left(E-\varepsilon_{k\beta\bar{\sigma}}-\varepsilon_{\sigma}+\varepsilon_{\bar{\sigma}}\right)\langle\langle c_{k\beta\bar{\sigma}}d_{\bar{\sigma}}^{+}d_{\sigma};d_{\sigma}^{+}\rangle\rangle_{E} = -\langle d_{\sigma}^{+}c_{k\beta\bar{\sigma}}\rangle + T_{k\beta\bar{\sigma}}\langle\langle d_{\sigma};d_{\sigma}^{+}\rangle\rangle_{E} \\ -T_{k\beta\bar{\sigma}}\langle\langle d_{\sigma}n_{\bar{\sigma}};d_{\sigma}^{+}\rangle\rangle_{E} + \sum_{q\alpha}\left[-T_{q\alpha\bar{\sigma}}\langle\langle c_{k\beta\bar{\sigma}}c_{q\alpha\bar{\sigma}}^{+}d_{\sigma};d_{\sigma}^{+}\rangle\rangle_{E} + T_{q\alpha\sigma}^{*}\langle\langle c_{k\beta\bar{\sigma}}d_{\bar{\sigma}}^{+}c_{q\alpha\sigma};d_{\sigma}^{+}\rangle\rangle_{E}\right]\end{aligned} \quad (17)$$

The consecutive application of the EOM method to these Green functions leads to the higher-order Green functions. In order to truncate higher correlation terms some adopted the standard Hartree-Fock truncation approximation[23] . That is why this approximation is useful to establish a set of equations for the first-order and second-order Green functions and solve easily them and obtain the corresponding self-energy $\Sigma^{r}$. However, this approximation is valid only for the relatively higher temperature region. On the other hand, the decoupling approximation worked out by Meir $et$ $al$[15,26,28] is sometimes used to establish a closed of equations. This decoupling approximation is based on assuming that the higher-oeder spin correlations of the electrode electrons can be neglected. This approximation describes sufficiently the phenomena not only above the Kondo temperature $T_K$, but also near $T_K$. However, it is not correct for much smaller temperature than $T_K$. To consider the whole regime of temperature including $T_K$, we adopt the decoupling approximation of Lacroix[29]. According to Ref. 29, the second-order Green functions can be written as the following forms .

$$\begin{aligned}\langle\langle c_{k\beta\sigma}d_{\bar{\sigma}}^{+}c_{q\alpha\bar{\sigma}};d_{\sigma}^{+}\rangle\rangle_{E} &\approx \langle d_{\bar{\sigma}}^{+}c_{q\alpha\bar{\sigma}}\rangle\langle\langle c_{k\beta\sigma};d_{\sigma}^{+}\rangle\rangle_{E} \\ \langle\langle c_{k\beta\sigma}c_{q\alpha\bar{\sigma}}^{+}d_{\bar{\sigma}};d_{\sigma}^{+}\rangle\rangle_{E} &\approx \langle c_{q\alpha\bar{\sigma}}^{+}d_{\bar{\sigma}}\rangle\langle\langle c_{k\beta\sigma};d_{\sigma}^{+}\rangle\rangle_{E} \\ \langle\langle c_{k\beta\bar{\sigma}}^{+}c_{q\alpha\sigma}d_{\bar{\sigma}};d_{\sigma}^{+}\rangle\rangle_{E} &\approx -\langle c_{k\beta\bar{\sigma}}^{+}d_{\bar{\sigma}}\rangle\langle\langle c_{q\alpha\sigma};d_{\sigma}^{+}\rangle\rangle_{E} \\ \langle\langle c_{k\beta\bar{\sigma}}^{+}d_{\sigma}c_{q\alpha\bar{\sigma}};d_{\sigma}^{+}\rangle\rangle_{E} &\approx -\langle c_{k\beta\bar{\sigma}}^{+}c_{q\alpha\bar{\sigma}}\rangle\langle\langle d_{\sigma};d_{\sigma}^{+}\rangle\rangle_{E} \\ \langle\langle c_{k\beta\bar{\sigma}}c_{q\alpha\bar{\sigma}}^{+}d_{\sigma};d_{\sigma}^{+}\rangle\rangle_{E} &\approx \langle c_{k\beta\bar{\sigma}}c_{q\alpha\bar{\sigma}}^{+}\rangle\langle\langle d_{\sigma};d_{\sigma}^{+}\rangle\rangle_{E} \\ \langle\langle c_{k\beta\bar{\sigma}}d_{\bar{\sigma}}^{+}c_{q\alpha\sigma};d_{\sigma}^{+}\rangle\rangle_{E} &\approx -\langle d_{\bar{\sigma}}^{+}c_{k\beta\bar{\sigma}}\rangle\langle\langle c_{q\alpha\sigma};d_{\sigma}^{+}\rangle\rangle_{E}\end{aligned} \quad (18)$$

First, applying the above mentioned approximation to Eq. (15) which is the equation of motion for $K_{k\beta\sigma,d\sigma}^{A(2)}(E)=\langle\langle c_{k\beta\sigma}n_{\bar{\sigma}};d_\sigma^+\rangle\rangle_E$ and considering the first term of Eq. (13), we can show that

$$\sum_{k\beta}\left[T_{k\beta\sigma}^*\langle\langle c_{k\beta\sigma}n_{\bar{\sigma}};d_\sigma^+\rangle\rangle_E^r\right]=\sum_{k\beta}|T_{k\beta\sigma}|^2 G_{\sigma\sigma}^{(2)}(E)g_{k\beta\sigma}^r(E)$$
$$+\sum_{k\beta}T_{k\beta\sigma}^*g_{k\beta\sigma}^r(E)\langle\langle c_{k\beta\sigma};d_\sigma^+\rangle\rangle_E^r\sum_{g\alpha}\left[T_{g\alpha\bar{\sigma}}^*\langle d_\sigma^+ c_{g\alpha\bar{\sigma}}\rangle - T_{g\alpha\bar{\sigma}}\langle c_{g\alpha\bar{\sigma}}^+ d_{\bar{\sigma}}\rangle\right]$$
(19)

The first term on the RHS of the above equation becomes $\Sigma_{0\sigma}^r(E)G_{\sigma\sigma}^{(2)}(E)$ by Eq. (12) and the second term becomes

$$\sum_{g\alpha}T_{g\alpha\bar{\sigma}}^*\langle d_\sigma^+ c_{g\alpha\bar{\sigma}}\rangle = -i\int\frac{d\varepsilon}{2\pi}\sum_{g\alpha}T_{g\alpha\bar{\sigma}}^*\langle\langle c_{k\beta\sigma};d_\sigma^+\rangle\rangle^<$$
$$= -i\int\frac{d\varepsilon}{2\pi}\sum_{g\alpha}|T_{g\alpha\bar{\sigma}}|^2\left[g_{g\alpha\bar{\sigma}}^r G_{\bar{\sigma}\bar{\sigma}}^< + g_{g\alpha\bar{\sigma}}^< G_{\bar{\sigma}\bar{\sigma}}^a\right]$$
(20)

with the help of the definition of the lesser Green function and its Fourier transformation equation. Remembering the definition of the 0-th self-energy and the half-width function, Eq. (20) can be written as the following.

$$-i\int\frac{d\varepsilon}{2\pi}\sum_{k\beta}|T_{k\beta\sigma}|^2[g_{k\beta\bar{\sigma}}^r G_{\bar{\sigma}\bar{\sigma}}^< + g_{k\beta\bar{\sigma}}^< G_{\bar{\sigma}\bar{\sigma}}^a]$$
$$= -i\int\frac{d\varepsilon}{2\pi}\sum_{k\beta}|T_{k\beta\sigma}|^2 g_{k\beta\bar{\sigma}}^r G_{\bar{\sigma}\bar{\sigma}}^< - i\int\frac{d\varepsilon}{2\pi}\sum_{k\beta}|T_{k\beta\sigma}|^2 2\pi if_\beta(\varepsilon)\delta(\varepsilon-\varepsilon_{k\beta\bar{\sigma}})G_{\bar{\sigma}\bar{\sigma}}^a(\varepsilon)$$
$$= -i\int\frac{d\varepsilon}{2\pi}\Sigma_{0\bar{\sigma}}^r G_{\bar{\sigma}\bar{\sigma}}^< + \int\frac{d\varepsilon}{2\pi}\sum_\beta\Gamma_{\bar{\sigma}}^\beta f_\beta(\varepsilon)G_{\bar{\sigma}\bar{\sigma}}^a(\varepsilon)$$

Again, with using Keldysh equation $G_{\sigma\sigma}^< = G_{\sigma\sigma}^r\Sigma_{0\sigma}^< G_{\sigma\sigma}^a$ and Ng *ansatz*, the lesser Green function becomes

$$G^< = \Sigma_0^<\left(\Sigma_0^r - \Sigma_0^a\right)^{-1}(G^r - G^a)$$
$$= i\sum_\beta f_\beta(\varepsilon)\Gamma_\sigma^\beta(-i\sum_\beta\Gamma_\sigma^\beta)^{-1}(G^r - G^a)$$
$$= -\frac{\Gamma_\sigma^L f_L + \Gamma_\sigma^R f_R}{\Gamma_\sigma^L + \Gamma_\sigma^R}(G_{0\sigma}^r - G_{0\sigma}^a)$$
$$= -\tilde{f}(\varepsilon)(G_{0\sigma}^r - G_{0\sigma}^a)$$
(21)

, where $\Sigma_0^r - \Sigma_0^a = -i\sum_\beta\Gamma_\sigma^\beta$ and $\Sigma_{0\sigma}^< = i\sum_\beta f_\beta(\varepsilon)\Gamma_\sigma^\beta$ are considered. The following function

$$\tilde{f}(\varepsilon) = \frac{\Gamma_\sigma^L f_L + \Gamma_\sigma^R f_R}{\Gamma^L + \Gamma^R}$$
(22)

can be considered as the renormalized Fermi distribution function due to the coupling with electrodes. This seems to be a weight-averaged distribution function with half-width functions.

Next, applying Eq. (18) to Eq. (16) which is the equation of motion for

$$K^{B(2)}_{k\beta\sigma,d\sigma}(E) = \langle\langle c^+_{k\beta\sigma} d_\sigma d_{\bar\sigma}; d^+_\sigma \rangle\rangle_E$$

and using the second term of Eq. (13) result in

$$\sum_{k\beta} T_{k\beta\bar\sigma} K^{B(2)r}_{k\beta\sigma,d\sigma}(E) = \sum_{k\beta} T_{k\beta\bar\sigma} \tilde{g}^r_{k\beta\bar\sigma}(E) \Big\{ -\langle c^+_{k\beta\bar\sigma} d_{\bar\sigma}\rangle + T^*_{k\beta\bar\sigma} G^{(2)r}_{\sigma\sigma} - \\ -\sum_{g\alpha} \Big[ T^*_{g\alpha\sigma} \langle c^+_{k\beta\bar\sigma} d_{\bar\sigma}\rangle K^r_{g\alpha\sigma,d\sigma} + T^*_{g\alpha\bar\sigma} \langle c^+_{k\beta\bar\sigma} c_{g\alpha\bar\sigma}\rangle G^r_{\sigma\sigma} \Big] \Big\} \quad (23)$$

, where the third term on the RHS becomes

$$\sum_{g\alpha} T^*_{g\alpha\sigma} \langle c^+_{k\beta\bar\sigma} d_{\bar\sigma}\rangle K^r_{g\alpha\sigma,d\sigma} = \langle c^+_{k\beta\bar\sigma} d_{\bar\sigma}\rangle \sum_{g\alpha} T^*_{g\alpha\sigma} K^r_{g\alpha\sigma,d\sigma} \\ = \langle c^+_{k\beta\bar\sigma} d_{\bar\sigma}\rangle \sum_{g\alpha} |T_{g\alpha\sigma}|^2 g^r_{g\alpha\sigma} G^r_{\sigma\sigma} = \langle c^+_{k\beta\bar\sigma} d_{\bar\sigma}\rangle \Sigma^r_{0\sigma} G^r_{\sigma\sigma} \quad (24)$$

Therefore Eq. (23) can be rewritten as follows.

$$\sum_{k\beta} T_{k\beta\bar\sigma} K^{B(2)r}_{k\beta\sigma,d\sigma}(E) = -(1 + \Sigma^r_{0\sigma} G^r_{\sigma\sigma}) \sum_{k\beta} T_{k\beta\bar\sigma} \tilde{g}^r_{k\beta\bar\sigma}(E) \langle c^+_{k\beta\bar\sigma} d_{\bar\sigma}\rangle \\ + \sum_{k\beta} |T_{k\beta\bar\sigma}|^2 \tilde{g}^r_{k\beta\bar\sigma}(E) G^{(2)r}_{\sigma\sigma} \\ - \sum_{k\beta}\sum_{g\alpha} T_{k\beta\bar\sigma} T^*_{g\alpha\bar\sigma} \tilde{g}^r_{k\beta\bar\sigma}(E) \langle c^+_{k\beta\bar\sigma} c_{g\alpha\bar\sigma}\rangle G^r_{\sigma\sigma} \quad (25)$$

, where $\tilde{g}^r_{k\beta\bar\sigma}(E) = (E - \tilde{\varepsilon}_{k\beta\bar\sigma} + i0^+)^{-1}$ and $\tilde{\varepsilon}_{k\beta\bar\sigma} = -\varepsilon_{k\beta\bar\sigma} + \varepsilon_\sigma + \varepsilon_{\bar\sigma} + U$.

Finally, applying Eq. (18) to Eq. (17) which is the equation of motion for

$$K^{C(2)}_{k\beta\sigma,d\sigma}(E) = \langle\langle c_{k\beta\bar\sigma} d^+_{\bar\sigma} d_\sigma; d^+_\sigma \rangle\rangle_E$$

and considering the third term of Eq. (13), the following equation is obtained

$$\sum_{k\beta} T^*_{k\beta\bar\sigma} K^{C(2)}_{k\beta\sigma,d\sigma}(E) = \sum_{k\beta} T^*_{k\beta\bar\sigma} \tilde{\tilde{g}}^r_{k\beta\bar\sigma}(E) \Big\{ -\langle d^+_\sigma c_{k\beta\bar\sigma}\rangle + T_{k\beta\bar\sigma} G^r_{\sigma\sigma} - \\ \sum_{g\alpha} \Big[ -T_{g\alpha\bar\sigma}(\delta_{k\beta,g\alpha} - \langle c^+_{g\alpha\bar\sigma} c_{k\beta\bar\sigma}\rangle) G^r_{\sigma\sigma} - T^*_{g\alpha\sigma} \langle d^+_\sigma c_{k\beta\bar\sigma}\rangle K^r_{g\alpha\sigma,d\sigma} \Big] \Big\} \quad (26)$$

and after tidying up we can rewrite the above equation as

$$\sum_{k\beta} T^*_{k\beta\bar\sigma} K^{C(2)r}_{k\beta\sigma,d\sigma}(E) = -(1 + \Sigma^r_{0\sigma} G^r_{\sigma\sigma}) \sum_{k\beta} T^*_{k\beta\bar\sigma} \tilde{\tilde{g}}^r_{k\beta\bar\sigma}(E) \langle d^+_\sigma c_{k\beta\bar\sigma}\rangle \\ - \sum_{k\beta} |T_{k\beta\bar\sigma}|^2 \tilde{\tilde{g}}^r_{k\beta\bar\sigma}(E) G^{(2)r}_{\sigma\sigma} \\ + \sum_{k\beta}\sum_{g\alpha} T^*_{k\beta\bar\sigma} T_{g\alpha\bar\sigma} \tilde{\tilde{g}}^r_{k\beta\bar\sigma}(E) \langle c^+_{g\alpha\bar\sigma} c_{k\beta\bar\sigma}\rangle G^r_{\sigma\sigma} \quad (27)$$

, where $\tilde{\tilde{g}}^r_{k\beta\bar\sigma} = (E - \tilde{\tilde{\varepsilon}}_{k\beta\bar\sigma} + i0^+)^{-1}$ and $\tilde{\tilde{\varepsilon}}_{k\beta\bar\sigma} = \varepsilon_{k\beta\bar\sigma} + \varepsilon_\sigma - \varepsilon_\sigma$.

From now on, we are going to enter the detailed calculations. We are going to calculate first the first terms of the RHS in Eq. (25) and Eq. (27) and then the other terms one after another.

1) The first term of Eq. (25)

$$\sum_{k\beta} T_{k\beta\bar{\sigma}} \tilde{g}^r_{k\beta\bar{\sigma}}(E) \langle c^+_{k\beta\bar{\sigma}} d_{\bar{\sigma}} \rangle = -i \int \frac{d\varepsilon}{2\pi} \sum_{k\beta} T_{k\beta\bar{\sigma}} \tilde{g}^r_{k\beta\bar{\sigma}}(E) \langle\langle d_{\bar{\sigma}}; c^+_{k\beta\bar{\sigma}} \rangle\rangle^<_E \quad (28)$$

must be obtained from the equation of motion for $K_{d\sigma,k\beta\sigma}(E) = \langle\langle d_{\sigma}; c^+_{k\beta\sigma} \rangle\rangle_E$, which using

$$E << A; B^+ >>_E = <\{A, B^+\}> - << A; [B^+, H]_- >>_E \quad (29)$$

and the analytic continuation of

$$(E - \varepsilon_{k\beta\sigma}) << d_\sigma; c^+_{k\beta\sigma} >>_E = T^*_{k\beta\sigma} G_{\sigma\sigma}(E)$$

emerged from Eq. (29), is related to

$$\langle\langle d_{\sigma}; c^+_{k\beta\sigma} \rangle\rangle^<_E = T^*_{k\beta\sigma} [g^{r(\varepsilon)}_{k\beta\sigma} G^{<(\varepsilon)}_{\sigma\sigma} + g^{<(\varepsilon)}_{k\beta\sigma} G^{a(\varepsilon)}_{\sigma\sigma}] \quad (30)$$

As a result, Eq. (28) leads to

$$\sum_{k\beta} T_{k\beta\bar{\sigma}} \tilde{g}^r_{k\beta\bar{\sigma}}(E) \langle c^+_{k\beta\bar{\sigma}} d_{\bar{\sigma}} \rangle = -i \int \frac{d\varepsilon}{2\pi} \sum_{k\beta} |T_{k\beta\bar{\sigma}}|^2 \tilde{g}^r_{k\beta\bar{\sigma}}(E) [g^{r(\varepsilon)}_{k\beta\bar{\sigma}} G^{<(\varepsilon)}_{\sigma\sigma} + g^{<(\varepsilon)}_{k\beta\bar{\sigma}} G^{a(\varepsilon)}_{\sigma\sigma}] \quad (31)$$

In order to calculate the first term of the RHS of the above equation, we use the following

$$g(\omega + i\eta) g(\omega' + i\eta') = \frac{g(\omega' + i\eta') + g(\omega + i\eta)}{\omega + \omega' + i\eta}$$

and then obtain the following equation.

$$\tilde{g}^r_{k\beta\bar{\sigma}}(E) g^r_{k\beta\bar{\sigma}}(E) =$$

$$= \frac{1}{E + \varepsilon_{k\beta\bar{\sigma}} - \varepsilon_\sigma - \varepsilon_{\bar{\sigma}} - U + i0^+} \frac{1}{\varepsilon - \varepsilon_{k\beta\bar{\sigma}} + i0^+} \quad (32)$$

$$= \frac{1}{E + \varepsilon - \varepsilon_\sigma - \varepsilon_{\bar{\sigma}} - U + i0^+} [\tilde{g}^r_{k\beta\bar{\sigma}}(E) + g^r_{k\beta\bar{\sigma}}(\varepsilon)]$$

Thus the first term of Eq. (31) can be written as

$$-i \int \frac{d\varepsilon}{2\pi} \sum_{k\beta} |T_{k\beta\bar{\sigma}}|^2 \tilde{g}^r_{k\beta\bar{\sigma}}(E) g^{r(\varepsilon)}_{k\beta\bar{\sigma}} G^{<(\varepsilon)}_{\sigma\sigma} =$$

$$= -i \int \frac{d\varepsilon}{2\pi} \frac{1}{E + \varepsilon - \varepsilon_\sigma - \varepsilon_{\bar{\sigma}} - U + i0^+} G^<_{\sigma\sigma}(\varepsilon) \sum_{k\beta} |T_{k\beta\bar{\sigma}}|^2 [\tilde{g}^r_{k\beta\bar{\sigma}}(E) + g^r_{k\beta\bar{\sigma}}(\varepsilon)] \quad (31\text{-}1)$$

$$= -i \int \frac{d\varepsilon}{2\pi} \frac{1}{E + \varepsilon - \varepsilon_\sigma - \varepsilon_{\bar{\sigma}} - U + i0^+} G^<_{\sigma\sigma}(\varepsilon) [\tilde{\Sigma}^r_{0\bar{\sigma}}(E) + \Sigma^r_{0\bar{\sigma}}(\varepsilon)]$$

, where

$$\tilde{\Sigma}^r_{0\bar{\sigma}}(E) = \sum_{k\beta} |T_{k\beta\bar{\sigma}}|^2 \tilde{g}^r_{k\beta\bar{\sigma}}(E)$$

$$= \sum_{k\beta} |T_{k\beta\bar{\sigma}}|^2 \frac{1}{E + \varepsilon_{k\beta\bar{\sigma}} - \varepsilon_\sigma - \varepsilon_{\bar{\sigma}} - U + i0^+}$$

$$= \sum_{\beta} \int \frac{d\varepsilon}{2\pi} \Gamma^\beta_{\bar{\sigma}} \frac{1}{E + \varepsilon - \varepsilon_\sigma - \varepsilon_{\bar{\sigma}} - U + i0^+}$$

In the same manner, we can rewrite the second term of (31).

$$-i\int \frac{d\varepsilon}{2\pi}\sum_{k\beta}|T_{k\beta\bar{\sigma}}|^2 \tilde{g}^r_{k\beta\bar{\sigma}}(E)g^<_{k\beta\sigma}(\varepsilon)G^a_{\sigma\sigma}(\varepsilon)=$$

$$=-i\int \frac{d\varepsilon}{2\pi}\sum_{k\beta}|T_{k\beta\bar{\sigma}}|^2 \frac{1}{E+\varepsilon_{k\beta\bar{\sigma}}-\varepsilon_\sigma-\varepsilon_{\bar{\sigma}}-U+i0^+}2\pi i f_\beta(\varepsilon)\delta(\varepsilon-\varepsilon_{k\beta\bar{\sigma}})G^a_{\sigma\sigma}(\varepsilon) \quad (31\text{-}2)$$

$$=\sum_\beta \int \frac{d\varepsilon}{2\pi}\Gamma^\beta_{\bar{\sigma}}\frac{1}{E+\varepsilon-\varepsilon_\sigma-\varepsilon_{\bar{\sigma}}-U+i\theta^+}f_\beta(\varepsilon)G^a_{\sigma\sigma}(\varepsilon)$$

Thus, Eq. (31) can be written as follows.

$$\sum_{k\beta}T_{k\beta\sigma}\tilde{g}^r_{k\beta\sigma}(E)\langle c^+_{k\beta\bar{\sigma}}d_{\bar{\sigma}}\rangle = \int \frac{d\varepsilon}{2\pi}\frac{1}{E+\varepsilon-\varepsilon_\sigma-\varepsilon_{\bar{\sigma}}-U+i\theta^+}$$
$$\left\{-iG^<_{\sigma\sigma}(\varepsilon)[\tilde{\Sigma}^r_{0\bar{\sigma}}(E)+\Sigma^r_{0\bar{\sigma}}(\varepsilon)]+\sum_\beta \Gamma^\beta_{\bar{\sigma}}f_\beta(\varepsilon)G^a_{\sigma\sigma}(\varepsilon)\right\} \quad (33)$$

With the help of Eq. (21), Eq. (31) becomes

$$\sum_{k\beta}T_{k\beta\sigma}\tilde{g}^r_{k\beta\sigma}(E)\langle c^+_{k\beta\bar{\sigma}}d_{\bar{\sigma}}\rangle = \int \frac{d\varepsilon}{2\pi}\frac{1}{E+\varepsilon-\varepsilon_\sigma-\varepsilon_{\bar{\sigma}}-U+i\theta^+}\frac{\sum_\beta \Gamma^\beta_{\bar{\sigma}}f_\beta(\varepsilon)}{\Gamma_{\bar{\sigma}}}$$
$$\left\{-i(G^r-G^a)[\tilde{\Sigma}^r_{0\bar{\sigma}}(E)+\Sigma^r_{0\bar{\sigma}}(\varepsilon)]+\Gamma_{\bar{\sigma}}G^a_{\sigma\sigma}(\varepsilon)\right\} \quad (34)$$

2) Now, we are going to calculate the first term of Eq. (27).

$$\sum_{k\beta}T^*_{k\beta\bar{\sigma}}\tilde{g}^r_{k\beta\bar{\sigma}}(E)\langle d^+_{\bar{\sigma}}c_{k\beta\bar{\sigma}}\rangle = -i\int \frac{d\varepsilon}{2\pi}\sum_{k\beta}T^*_{k\beta\bar{\sigma}}\tilde{g}^r_{k\beta\bar{\sigma}}(E)K^<_{k\beta\bar{\sigma},d\bar{\sigma}}(\varepsilon)$$
$$=-i\int \frac{d\varepsilon}{2\pi}\sum_{k\beta}|T_{k\beta\bar{\sigma}}|^2 \tilde{g}^r_{k\beta\bar{\sigma}}(E)[g^r_{k\beta\bar{\sigma}}(\varepsilon)G^<_{\sigma\sigma}(\varepsilon)+g^<_{k\beta\bar{\sigma}}(\varepsilon)G^a_{\sigma\sigma}(\varepsilon)] \quad (35)$$

With the help of

$$\tilde{g}^r_{k\beta\bar{\sigma}}(E)g^r_{k\beta\bar{\sigma}}(\varepsilon) = \frac{1}{E-\varepsilon_{k\beta\bar{\sigma}}-\varepsilon_\sigma+\varepsilon_{\bar{\sigma}}+i0^+}\frac{1}{\varepsilon-\varepsilon_{k\beta\bar{\sigma}}+i0^+}$$
$$=\frac{1}{E-\varepsilon-\varepsilon_\sigma+\varepsilon_{\bar{\sigma}}+i0^+}[g^r_{k\beta\bar{\sigma}}(\varepsilon)-\tilde{g}^r_{k\beta\bar{\sigma}}(E)] \quad (36)$$

, the first term on the RHS of Eq. (38) can be written as follows

$$-i\int \frac{d\varepsilon}{2\pi}\sum_{k\beta}|T_{k\beta\bar{\sigma}}|^2 \tilde{g}^r_{k\beta\bar{\sigma}}(E)g^r_{k\beta\bar{\sigma}}(\varepsilon)G^<_{\sigma\sigma}(\varepsilon)$$
$$=-i\int \frac{d\varepsilon}{2\pi}\sum_{k\beta}|T_{k\beta\bar{\sigma}}|^2 \frac{1}{E-\varepsilon-\varepsilon_\sigma+\varepsilon_{\bar{\sigma}}+i0^+}[g^r_{k\beta\bar{\sigma}}(\varepsilon)-\tilde{g}^r_{k\beta\bar{\sigma}}(E)]G^<_{\sigma\sigma}(\varepsilon)$$
$$=-i\int \frac{d\varepsilon}{2\pi}\frac{1}{E-\varepsilon-\varepsilon_\sigma+\varepsilon_{\bar{\sigma}}+i0^+}G^<_{\sigma\sigma}(\varepsilon)\sum_{k\beta}|T_{k\beta\bar{\sigma}}|^2[g^r_{k\beta\bar{\sigma}}(\varepsilon)-\tilde{g}^r_{k\beta\bar{\sigma}}(E)] \quad (35\text{-}1)$$
$$=-i\int \frac{d\varepsilon}{2\pi}\frac{1}{E-\varepsilon-\varepsilon_\sigma+\varepsilon_{\bar{\sigma}}+i0^+}G^<_{\sigma\sigma}(\varepsilon)[\Sigma^r_{0\bar{\sigma}}(\varepsilon)-\tilde{\Sigma}^r_{0\bar{\sigma}}(E)]$$

, where

$$\tilde{\tilde{\Sigma}}^r_{0\bar{\sigma}}(E) = \sum_{k\beta} |T_{k\beta\bar{\sigma}}|^2 \tilde{\tilde{g}}^r_{k\beta\bar{\sigma}}(E)$$

$$= \sum_{k\beta} |T_{k\beta\bar{\sigma}}|^2 \frac{1}{E - \varepsilon_{k\beta\bar{\sigma}} - \varepsilon_\sigma + \varepsilon_{\bar{\sigma}} + i0^+}$$

$$= \sum_{\beta} \int \frac{d\varepsilon}{2\pi} \Gamma^\beta_{\bar{\sigma}} \frac{1}{E - \varepsilon - \varepsilon_\sigma + \varepsilon_{\bar{\sigma}} + i0^+}$$

And the second term on the RHS of Eq. (38) becomes.

$$-i \int \frac{d\varepsilon}{2\pi} \sum_{k\beta} |T_{k\beta\bar{\sigma}}|^2 \tilde{\tilde{g}}^r_{k\beta\bar{\sigma}}(E) g^<_{k\beta\bar{\sigma}}(\varepsilon) G^a_{\bar{\sigma}\bar{\sigma}}(\varepsilon) =$$

$$= -i \int \frac{d\varepsilon}{2\pi} \sum_{k\beta} |T_{k\beta\bar{\sigma}}|^2 \frac{1}{E - \varepsilon_{k\beta\bar{\sigma}} - \varepsilon_\sigma + \varepsilon_{\bar{\sigma}} + i0^+} 2\pi i f_\beta(\varepsilon) \delta(\varepsilon - \varepsilon_{k\beta\sigma}) G^a_{\bar{\sigma}\bar{\sigma}}(\varepsilon) \qquad (35\text{-}2)$$

$$= \sum_{\beta} \int \frac{d\varepsilon}{2\pi} \Gamma^\beta_{\bar{\sigma}} \frac{1}{E - \varepsilon - \varepsilon_\sigma + \varepsilon_{\bar{\sigma}} + i0^+} f_\beta(\varepsilon) G^a_{\bar{\sigma}\bar{\sigma}}(\varepsilon)$$

From Eq. (35-1) and (35-2), Eq. (35) becomes

$$\sum_{k\beta} T^*_{k\beta\bar{\sigma}} \tilde{\tilde{g}}^r_{k\beta\bar{\sigma}}(E) \langle d^+_{\bar{\sigma}} c_{k\beta\bar{\sigma}} \rangle = \int \frac{d\varepsilon}{2\pi} \frac{1}{E - \varepsilon - \varepsilon_\sigma + \varepsilon_{\bar{\sigma}} + i0^+} \left\{ -iG^<_{\bar{\sigma}\bar{\sigma}}(\varepsilon) \left[ \Sigma^r_{o\bar{\sigma}}(\varepsilon) - \tilde{\tilde{\Sigma}}^r_{o\bar{\sigma}}(\varepsilon) \right] + \sum_\beta \Gamma^\beta_{\bar{\sigma}} f_\beta(\varepsilon) G^a_{\bar{\sigma}\bar{\sigma}}(\varepsilon) \right\}$$

$$= \int \frac{d\varepsilon}{2\pi} \frac{1}{E - \varepsilon - \varepsilon_\sigma + \varepsilon_{\bar{\sigma}} + i0^+} \frac{\sum_\beta \Gamma^\beta_{\bar{\sigma}} f_\beta(\varepsilon)}{\Gamma_{\bar{\sigma}}} \times$$

$$\times \left\{ i(G^r(\varepsilon) - G^a(\varepsilon)) \left[ \Sigma^r_{o\bar{\sigma}}(\varepsilon) - \tilde{\tilde{\Sigma}}^r_{o\bar{\sigma}}(\varepsilon) \right] + \Gamma_{\bar{\sigma}} G^a_{\bar{\sigma}\bar{\sigma}}(\varepsilon) \right\}$$

(37)

3) Now, we consider the third term of Eq. (25) $-\sum\sum T_{k\beta\bar{\sigma}} T^*_{g\alpha\bar{\sigma}} \tilde{g}^r_{k\beta\bar{\sigma}}(E) \langle c^+_{k\beta\bar{\sigma}} c_{g\alpha\bar{\sigma}} \rangle$.

To do this, we must first look for the equation of motion for the electrode-electrode

$$M_{g\alpha\sigma,k\beta\sigma}(\varepsilon) = \langle\langle c_{g\alpha\sigma}; c^+_{k\beta\sigma} \rangle\rangle_\varepsilon$$

Using the following relations

$$\langle \{c_{g\alpha\sigma}, c^+_{k\beta\sigma}\} \rangle = \delta_{g\alpha,k\beta}$$

$$[c_{g\alpha\sigma}, H] = \varepsilon_{g\alpha\sigma} c_{g\alpha\sigma} + T_{g\alpha\sigma} d_\sigma$$

gives the equation of motion for $M_{g\alpha\sigma,k\beta\sigma}(\varepsilon) = \langle\langle c_{g\alpha\sigma}; c^+_{k\beta\sigma} \rangle\rangle_\varepsilon$.

$$(\varepsilon - \varepsilon_{g\alpha\sigma}) \langle\langle c_{g\alpha\sigma}; c^+_{k\beta\sigma} \rangle\rangle^r_\varepsilon = \delta_{g\alpha,k\beta} + T_{g\alpha\sigma} \langle\langle d_\sigma; c^+_{k\beta\sigma} \rangle\rangle^r_\varepsilon \qquad (38)$$

With the help of the equation of motion for $\langle\langle d_\sigma; c^+_{k\beta\sigma} \rangle\rangle_\varepsilon$, the above given equation results in

$$(\varepsilon - \varepsilon_{g\alpha\sigma}) M^r_{g\alpha\sigma,k\beta\sigma}(\varepsilon) = \delta_{g\alpha,k\beta} + T_{g\alpha\sigma} T^*_{k\beta\sigma} [g^r_{k\beta\sigma}(\varepsilon) G^r_{\sigma\sigma}(\varepsilon)] \qquad (39)$$

$$M^r_{g\alpha\sigma,k\beta\sigma}(\varepsilon) = g^r_{g\alpha\sigma} \delta_{g\alpha,k\beta} + T_{g\alpha\sigma} T^*_{k\beta\sigma} g^r_{g\alpha\sigma}(\varepsilon) g^r_{k\beta\sigma}(\varepsilon) G^r_{\sigma\sigma}(\varepsilon) \qquad (40)$$

In addition, taking into account the rule for the following analytic continuation

$$(ABC)^< = (AB)^r C^< + (AB)^< C^a = A^r B^r C^< + A^r B^< C^a + A^< B^a C^a$$

leads to
$$M^<_{g\alpha\sigma,k\beta\sigma}(\varepsilon) = g^r_{g\alpha\sigma}\delta_{g\alpha,k\beta} + $$
$$+ T_{g\alpha\sigma}T^*_{k\beta\sigma}[g^r_{g\alpha\sigma}(\varepsilon)g^r_{k\beta\sigma}(\varepsilon)G^<_{\sigma\sigma}(\varepsilon) + g^r_{g\alpha\sigma}(\varepsilon)g^<_{k\beta\sigma}(\varepsilon)G^a_{\sigma\sigma}(\varepsilon) + g^<_{g\alpha\sigma}(\varepsilon)g^a_{k\beta\sigma}(\varepsilon)G^a_{\sigma\sigma}(\varepsilon)]$$
(41)

Thus the third term of Eq. (25) becomes
$$-\sum_{k\beta}\sum_{g\alpha}T_{k\beta\bar\sigma}T^*_{g\alpha\bar\sigma}\tilde g^r_{k\beta\bar\sigma}(E)\langle c^+_{k\beta\bar\sigma}c_{g\alpha\bar\sigma}\rangle$$
$$= -i\int\frac{d\varepsilon}{2\pi}\Bigg[\sum_{k\beta}\sum_{g\alpha}g^<_{g\alpha\bar\sigma}(\varepsilon)\delta_{g\alpha,k\beta}\tilde g^r_{k\beta\bar\sigma}(E)T_{k\beta\bar\sigma}T^*_{g\alpha\bar\sigma} +$$
$$+\sum_{k\beta}\sum_{g\alpha}|T_{k\beta\bar\sigma}|^2|T_{g\alpha\bar\sigma}|^2\tilde g^r_{k\beta\bar\sigma}(E)[g^r_{g\alpha\bar\sigma}(\varepsilon)g^r_{k\beta\bar\sigma}(\varepsilon)G^<_{\bar\sigma\bar\sigma}(\varepsilon) + g^r_{g\alpha\bar\sigma}(\varepsilon)g^<_{k\beta\bar\sigma}(\varepsilon)G^a_{\bar\sigma\bar\sigma}(\varepsilon) + g^<_{g\alpha\bar\sigma}(\varepsilon)g^a_{k\beta\bar\sigma}(\varepsilon)G^a_{\bar\sigma\bar\sigma}(\varepsilon)]\Bigg]$$
(42)

, where the first term on the RHS is
$$\sum_{k\beta}\sum_{g\alpha}T_{k\beta\bar\sigma}T^*_{g\alpha\bar\sigma}\tilde g^r_{k\beta\bar\sigma}(E)(-i\int\frac{d\varepsilon}{2\pi}g^<_{g\alpha\bar\sigma}(\varepsilon))\delta_{g\alpha,k\beta}$$
$$= \sum_{k\beta}|T_{k\beta\bar\sigma}|^2\tilde g^r_{k\beta\bar\sigma}(E)[-i\int\frac{d\varepsilon}{2\pi}2\pi i f_\beta(\varepsilon)\delta(\varepsilon - \varepsilon_{k\beta\bar\sigma})]$$
$$= \int\frac{d\varepsilon}{2\pi}\sum_{k\beta}|T_{k\beta\bar\sigma}|^2\tilde g^r_{k\beta\bar\sigma}(E)2\pi f_\beta(\varepsilon)\delta(\varepsilon-\varepsilon_{k\beta\bar\sigma})$$
$$= \sum_\beta\int\frac{d\varepsilon}{2\pi}\Gamma^\beta_{\bar\sigma}\frac{1}{E+\varepsilon-\varepsilon_\sigma-\varepsilon_{\bar\sigma}-U+i0^+}f_\beta(\varepsilon)$$
(42-1)

and the second is
$$-i\int\frac{d\varepsilon}{2\pi}\sum_{k\beta}\sum_{g\alpha}|T_{k\beta\bar\sigma}|^2|T_{g\alpha\bar\sigma}|^2\tilde g^r_{k\beta\bar\sigma}(E)g^r_{g\alpha\bar\sigma}(\varepsilon)g^r_{k\beta\bar\sigma}(\varepsilon)G^<_{\bar\sigma\bar\sigma}(\varepsilon)$$
$$= -i\int\frac{d\varepsilon}{2\pi}\Sigma^r_{0\bar\sigma}(\varepsilon)\sum_{k\beta}|T_{k\beta\bar\sigma}|^2\tilde g^r_{k\beta\bar\sigma}(E)g^r_{k\beta\bar\sigma}(\varepsilon)G^<_{\bar\sigma\bar\sigma}(\varepsilon)$$

According to Eq. (32), the above equation can be written as follows.
$$-i\int\frac{d\varepsilon}{2\pi}\Sigma^r_{0\bar\sigma}(\varepsilon)\frac{1}{E+\varepsilon-\varepsilon_\sigma-\varepsilon_{\bar\sigma}-U+i0^+}G^<_{\bar\sigma\bar\sigma}(\varepsilon)\sum_{k\beta}|T_{k\beta\bar\sigma}|^2[\tilde g^r_{k\beta\bar\sigma}(E)+g^r_{k\beta\bar\sigma}(\varepsilon)]$$
$$= -i\int\frac{d\varepsilon}{2\pi}\Sigma^r_{0\bar\sigma}(\varepsilon)\frac{1}{E+\varepsilon-\varepsilon_\sigma-\varepsilon_{\bar\sigma}-U+i0^+}G^<_{\bar\sigma\bar\sigma}(\varepsilon)[\tilde\Sigma^r_{0\bar\sigma}(E)+\Sigma^r_{0\bar\sigma}(\varepsilon)]$$
(42-2)

Similarly, the third is as follows
$$-i\int\frac{d\varepsilon}{2\pi}\sum_{k\beta}\sum_{g\alpha}|T_{k\beta\bar\sigma}|^2|T_{g\alpha\bar\sigma}|^2\tilde g^r_{k\beta\bar\sigma}(E)g^r_{g\alpha\bar\sigma}(\varepsilon)g^<_{k\beta\bar\sigma}(\varepsilon)G^a_{\bar\sigma\bar\sigma}(\varepsilon)$$
$$= -i\int\frac{d\varepsilon}{2\pi}\Sigma^r_{0\bar\sigma}(\varepsilon)\sum_{k\beta}|T_{k\beta\bar\sigma}|^2\tilde g^r_{k\beta\bar\sigma}(E)2\pi i f_\beta(\varepsilon)\delta(\varepsilon-\varepsilon_{k\beta\bar\sigma})G^a_{\bar\sigma\bar\sigma}(\varepsilon)$$
(42-3)
$$= \int\frac{d\varepsilon}{2\pi}\Sigma^r_{0\bar\sigma}(\varepsilon)\sum_\beta\Gamma^\beta_{\bar\sigma}\frac{1}{E+\varepsilon-\varepsilon_\sigma-\varepsilon_{\bar\sigma}-U+i0^+}f_\beta(\varepsilon)G^a_{\bar\sigma\bar\sigma}(\varepsilon)$$

and the fourth is as follows.

$$-i\int \frac{d\varepsilon}{2\pi} \sum_{k\beta} \sum_{g\alpha} |T_{k\beta\bar{\sigma}}|^2 |T_{g\alpha\bar{\sigma}}|^2 \tilde{g}^r_{k\beta\bar{\sigma}}(E) g^<_{g\alpha\bar{\sigma}}(\varepsilon) g^a_{k\beta\bar{\sigma}}(\varepsilon) G^a_{\bar{\sigma}\bar{\sigma}}(\varepsilon)$$

$$=-i\int \frac{d\varepsilon}{2\pi} \sum_{k\beta} \sum_{g\alpha} |T_{k\beta\bar{\sigma}}|^2 |T_{g\alpha\bar{\sigma}}|^2 \tilde{g}^r_{k\beta\bar{\sigma}}(E) 2\pi i f_\alpha(\varepsilon) \delta(\varepsilon-\varepsilon_{g\alpha\bar{\sigma}}) g^a_{k\beta\bar{\sigma}}(\varepsilon) G^a_{\bar{\sigma}\bar{\sigma}}(\varepsilon)$$

$$=\int \frac{d\varepsilon}{2\pi} \sum_\alpha \Gamma^\alpha_{\bar{\sigma}} f_\alpha(\varepsilon) \sum_{k\beta} |T_{k\beta\bar{\sigma}}|^2 \tilde{g}^r_{k\beta\bar{\sigma}}(E) g^a_{k\beta\bar{\sigma}}(\varepsilon) G^a_{\bar{\sigma}\bar{\sigma}}(\varepsilon)$$

Also, the following relation is established

$$\tilde{g}^r_{k\beta\bar{\sigma}}(E) g^a_{k\beta\bar{\sigma}}(\varepsilon) = \frac{1}{E+\varepsilon-\varepsilon_\sigma-\varepsilon_{\bar{\sigma}}-U+i0^+}[\tilde{g}^r_{k\beta\bar{\sigma}}(E)+g^a_{k\beta\bar{\sigma}}(\varepsilon)]$$

, similar to Eq. (32). So the fourth term of Eq. (48) becomes

$$\int \frac{d\varepsilon}{2\pi} \frac{1}{E+\varepsilon-\varepsilon_\sigma-\varepsilon_{\bar{\sigma}}-U+i0^+} G^a_{\bar{\sigma}\bar{\sigma}}(\varepsilon) \sum_\alpha \Gamma^\alpha_{\bar{\sigma}} f_\alpha(\varepsilon) \sum_{k\beta} |T_{k\beta\bar{\sigma}}|^2 [\tilde{g}^r_{k\beta\bar{\sigma}}(E)+g^a_{k\beta\bar{\sigma}}(\varepsilon)]$$

$$=\int \frac{d\varepsilon}{2\pi} \frac{1}{E+\varepsilon-\varepsilon_\sigma-\varepsilon_{\bar{\sigma}}-U+i0^+} G^a_{\bar{\sigma}\bar{\sigma}}(\varepsilon) \sum_\alpha \Gamma^\alpha_{\bar{\sigma}} f_\alpha(\varepsilon) [\tilde{\Sigma}^r_{0\bar{\sigma}}(E)+\Sigma^r_{0\bar{\sigma}}(\varepsilon)]$$

(42-4)

At last, the third term of Eq. (25) results in the following.

$$-\sum_{k\beta} \sum_{g\alpha} T_{k\beta\bar{\sigma}} T^*_{g\alpha\bar{\sigma}} \tilde{g}^r_{k\beta\bar{\sigma}}(E) \langle c^+_{k\beta\bar{\sigma}} c_{g\alpha\bar{\sigma}} \rangle$$

$$=\int \frac{d\varepsilon}{2\pi} \frac{1}{E+\varepsilon-\varepsilon_\sigma-\varepsilon_{\bar{\sigma}}-U+i0^+} \left[(-i)\Sigma^r_{o\bar{\sigma}}(\varepsilon) G^<_{\bar{\sigma}\bar{\sigma}}(\varepsilon)[\Sigma^r_{o\bar{\sigma}}(E)+\Sigma^r_{o\bar{\sigma}}(\varepsilon)] \right.$$

$$\left. +\left[\sum_\beta \Gamma^\beta_{\bar{\sigma}} f_\beta(\varepsilon)\right] \times \{1+G^a_{\bar{\sigma}\bar{\sigma}}(\varepsilon)[\Sigma^r_{o\bar{\sigma}}(E)+\Sigma^r_{o\bar{\sigma}}(\varepsilon)+\Sigma^a_{o\bar{\sigma}}(\varepsilon)]\}\right]$$

(43)

4) The final turn is the calculation of the third term of Eq. (27).

As in 3), using the equation of motion for electrode-electrode Green function

$$M_{k\beta\bar{\sigma},g\alpha\bar{\sigma}}(\varepsilon) =<< c_{k\beta\bar{\sigma}}; c^+_{g\alpha\bar{\sigma}} >>_\varepsilon$$

given by

$$(\varepsilon-\varepsilon_{k\beta\bar{\sigma}}) M_{k\beta\bar{\sigma},g\alpha\bar{\sigma}}(\varepsilon) = \delta_{g\alpha,k\beta} + T_{k\beta\bar{\sigma}} T^*_{g\alpha\bar{\sigma}} [g^r_{g\alpha\bar{\sigma}}(\varepsilon) G^r_{\bar{\sigma}\bar{\sigma}}(\varepsilon)]$$ (44)

$$M^r_{k\beta\bar{\sigma},g\alpha\bar{\sigma}}(\varepsilon) = g^r_{k\beta\bar{\sigma}} \delta_{g\alpha,k\beta} + T_{k\beta\bar{\sigma}} T^*_{g\alpha\bar{\sigma}} g^r_{k\beta\bar{\sigma}}(\varepsilon) g^r_{g\alpha\bar{\sigma}}(\varepsilon) G^r_{\bar{\sigma}\bar{\sigma}}(\varepsilon)$$ (45)

, the electrode-electrode lesser Green function $M^<_{k\beta\bar{\sigma},g\alpha\bar{\sigma}}(\varepsilon)$ has the form of the following.

$$M^<_{k\beta\bar{\sigma},g\alpha\bar{\sigma}}(\varepsilon) = g^<_{k\beta\bar{\sigma}}(\varepsilon) \delta_{k\beta,g\alpha} +$$
$$+T_{k\beta\bar{\sigma}} T^*_{g\alpha\bar{\sigma}} [g^r_{k\beta\bar{\sigma}}(\varepsilon) g^<_{g\alpha\bar{\sigma}}(\varepsilon) G^a_{\bar{\sigma}\bar{\sigma}}(\varepsilon) + g^r_{k\beta\bar{\sigma}}(\varepsilon) g^<_{g\alpha\bar{\sigma}}(\varepsilon) G^a_{\bar{\sigma}\bar{\sigma}}(\varepsilon) + g^<_{k\beta\bar{\sigma}}(\varepsilon) g^a_{g\alpha\bar{\sigma}}(\varepsilon) G^a_{\bar{\sigma}\bar{\sigma}}(\varepsilon)]$$

(46)

Considering the above equation in the third term of Eq. (27) yields the following.

$$\sum_{k\beta}\sum_{g\alpha}T^*_{k\beta\bar{\sigma}}T_{g\alpha\bar{\sigma}}\tilde{\tilde{g}}^r_{k\beta\bar{\sigma}}(E)\langle c^+_{g\alpha\bar{\sigma}}c_{k\beta\bar{\sigma}}\rangle = -i\int\frac{d\varepsilon}{2\pi}\sum_{k\beta}\sum_{g\alpha}T^*_{k\beta\bar{\sigma}}T_{g\alpha\bar{\sigma}}\tilde{\tilde{g}}^r_{k\beta\bar{\sigma}}(E)\times$$
$$\times\{g^<_{k\beta\bar{\sigma}}(\varepsilon)\delta_{k\beta,g\alpha} + T_{k\beta\bar{\sigma}}T^*_{g\alpha\bar{\sigma}}[g^r_{k\beta\bar{\sigma}}(\varepsilon)g^r_{g\alpha\bar{\sigma}}(\varepsilon)G^<_{\bar{\sigma}\bar{\sigma}}(\varepsilon) + g^r_{k\beta\bar{\sigma}}(\varepsilon)g^<_{g\alpha\bar{\sigma}}(\varepsilon)G^a_{\bar{\sigma}\bar{\sigma}}(\varepsilon) + g^<_{k\beta\bar{\sigma}}(\varepsilon)g^a_{g\alpha\bar{\sigma}}(\varepsilon)G^a_{\bar{\sigma}\bar{\sigma}}(\varepsilon)]\}$$
(47)

Also, as in 3), the first term of Eq.( 47) is

$$-i\int\frac{d\varepsilon}{2\pi}\sum_{k\beta}\sum_{g\alpha}T^*_{k\beta\bar{\sigma}}T_{g\alpha\bar{\sigma}}\tilde{\tilde{g}}^r_{k\beta\bar{\sigma}}(E)g^<_{k\beta\bar{\sigma}}(\varepsilon)\delta_{k\beta,g\alpha}$$
$$= -i\int\frac{d\varepsilon}{2\pi}\sum_{k\beta}|T_{k\beta\bar{\sigma}}|^2\tilde{\tilde{g}}^r_{k\beta\bar{\sigma}}(E)2\pi i f_\beta(\varepsilon)\delta(\varepsilon - \varepsilon_{k\beta\bar{\sigma}}) \quad (47\text{-}1)$$
$$= \sum_\beta\int\frac{d\varepsilon}{2\pi}\Gamma^\beta_{\bar{\sigma}}\frac{1}{E-\varepsilon-\varepsilon_\sigma+\varepsilon_{\bar{\sigma}}+i0^+}f_\beta(\varepsilon)$$

and the second is

$$-i\int\frac{d\varepsilon}{2\pi}\sum_{k\beta}\sum_{g\alpha}|T_{k\beta\bar{\sigma}}|^2|T_{g\alpha\bar{\sigma}}|^2\tilde{\tilde{g}}^r_{k\beta\bar{\sigma}}(E)g^r_{k\beta\bar{\sigma}}(\varepsilon)g^r_{g\alpha\bar{\sigma}}(\varepsilon)G^<_{\bar{\sigma}\bar{\sigma}}(\varepsilon)$$
$$= -i\int\frac{d\varepsilon}{2\pi}\Sigma^r_{0\bar{\sigma}}(\varepsilon)\sum_{k\beta}|T_{k\beta\bar{\sigma}}|^2\tilde{\tilde{g}}^r_{k\beta\bar{\sigma}}(E)g^r_{k\beta\bar{\sigma}}(\varepsilon)G^<_{\bar{\sigma}\bar{\sigma}}(\varepsilon)$$
(47-2-a)

Using the relation given by

$$\tilde{\tilde{g}}^r_{k\beta\bar{\sigma}}(E)g^r_{k\beta\bar{\sigma}}(\varepsilon) = \frac{1}{E-\varepsilon-\varepsilon_\sigma+\varepsilon_{\bar{\sigma}}+i0^+}[g^r_{k\beta\bar{\sigma}}(\varepsilon) - \tilde{\tilde{g}}^r_{k\beta\bar{\sigma}}(E)]$$

, Eq. (47-2-a) becomes

$$-i\int\frac{d\varepsilon}{2\pi}\Sigma^r_{0\bar{\sigma}}(\varepsilon)G^<_{\bar{\sigma}\bar{\sigma}}(\varepsilon)\sum_{k\beta}|T_{k\beta\bar{\sigma}}|^2[g^r_{k\beta\bar{\sigma}}(\varepsilon) - \tilde{\tilde{g}}^r_{k\beta\bar{\sigma}}(E)]$$
$$= -i\int\frac{d\varepsilon}{2\pi}\Sigma^r_{0\bar{\sigma}}(\varepsilon)\frac{1}{E-\varepsilon-\varepsilon_\sigma+\varepsilon_{\bar{\sigma}}+i0^+}G^<_{\bar{\sigma}\bar{\sigma}}(\varepsilon)[\Sigma^r_{0\bar{\sigma}}(\varepsilon) - \tilde{\tilde{\Sigma}}^r_{0\bar{\sigma}}(E)]$$
(47-2-b)

The third is

$$-i\int\frac{d\varepsilon}{2\pi}\sum_{k\beta}\sum_{g\alpha}|T_{k\beta\bar{\sigma}}|^2|T_{g\alpha\bar{\sigma}}|^2\tilde{\tilde{g}}^r_{k\beta\bar{\sigma}}(E)g^r_{k\beta\bar{\sigma}}(\varepsilon)g^<_{g\alpha\bar{\sigma}}(\varepsilon)G^a_{\bar{\sigma}\bar{\sigma}}(\varepsilon)$$
$$= -i\int\frac{d\varepsilon}{2\pi}\sum_{k\beta}\sum_{g\alpha}|T_{k\beta\bar{\sigma}}|^2|T_{g\alpha\bar{\sigma}}|^2\tilde{\tilde{g}}^r_{k\beta\bar{\sigma}}(E)g^r_{k\beta\bar{\sigma}}(\varepsilon)2\pi i f_\alpha(\varepsilon)\delta(\varepsilon-\varepsilon_{g\alpha\bar{\sigma}})G^a_{\bar{\sigma}\bar{\sigma}}(\varepsilon)$$
$$= \int\frac{d\varepsilon}{2\pi}G^a_{\bar{\sigma}\bar{\sigma}}(\varepsilon)[\sum_\alpha\Gamma^\alpha_{\bar{\sigma}}f_\alpha(\varepsilon)]\sum_{k\beta}|T_{k\beta\bar{\sigma}}|^2\tilde{\tilde{g}}^r_{k\beta\bar{\sigma}}(E)g^r_{k\beta\bar{\sigma}}(\varepsilon) \quad (47\text{-}3)$$
$$= \int\frac{d\varepsilon}{2\pi}[\sum_\alpha\Gamma^\alpha_{\bar{\sigma}}f_\alpha(\varepsilon)]\frac{1}{E-\varepsilon-\varepsilon_\sigma+\varepsilon_{\bar{\sigma}}+i0^+}G^a_{\bar{\sigma}\bar{\sigma}}(\varepsilon)\sum_{k\beta}|T_{k\beta\bar{\sigma}}|^2[g^r_{k\beta\bar{\sigma}}(\varepsilon) - \tilde{\tilde{g}}^r_{k\beta\bar{\sigma}}(E)]$$
$$= \int\frac{d\varepsilon}{2\pi}[\sum_\alpha\Gamma^\alpha_{\bar{\sigma}}f_\alpha(\varepsilon)]\frac{1}{E-\varepsilon-\varepsilon_\sigma+\varepsilon_{\bar{\sigma}}+i0^+}G^a_{\bar{\sigma}\bar{\sigma}}(\varepsilon)[\Sigma^r_{0\bar{\sigma}}(\varepsilon) - \tilde{\tilde{\Sigma}}^r_{0\bar{\sigma}}(E)]$$

and the fourth is

$$-i\int\frac{d\varepsilon}{2\pi}\sum_{k\beta}\sum_{g\alpha}\left|T_{k\beta\bar{\sigma}}\right|^{2}\left|T_{g\alpha\bar{\sigma}}\right|^{2}\tilde{g}_{k\beta\bar{\sigma}}^{r}(E)g_{k\beta\bar{\sigma}}^{<}(\varepsilon)g_{g\alpha\bar{\sigma}}^{a}(\varepsilon)G_{\bar{\sigma}\bar{\sigma}}^{a}(\varepsilon)$$

$$=-i\int\frac{d\varepsilon}{2\pi}\Sigma_{0\bar{\sigma}}^{a}(\varepsilon)\sum_{k\beta}\left|T_{k\beta\bar{\sigma}}\right|^{2}\tilde{g}_{k\beta\bar{\sigma}}^{r}(E)2\pi if_{\beta}(\varepsilon)\delta(\varepsilon-\varepsilon_{k\beta\bar{\sigma}})G_{\bar{\sigma}\bar{\sigma}}^{a}(\varepsilon) \qquad (47\text{-}4)$$

$$=-i\int\frac{d\varepsilon}{2\pi}\Sigma_{0\bar{\sigma}}^{a}(\varepsilon)[\sum_{\beta}\Gamma_{\bar{\sigma}}^{\beta}f_{\beta}(\varepsilon)]\frac{1}{E-\varepsilon-\varepsilon_{\sigma}+\varepsilon_{\bar{\sigma}}+i0^{+}}G_{\bar{\sigma}\bar{\sigma}}^{a}(\varepsilon)$$

Putting the equations given in 1)~4) into Eq. (25) and (27), the following equation of motion for the second Green function $G_{\sigma\sigma}^{(2)r}(E)$ can be obtained.

$$(E-E_{\sigma}-U-\Sigma_{03\sigma}^{r})G_{\sigma\sigma}^{(2)r}=\langle n_{\bar{\sigma}}\rangle-(\tilde{P}_{\sigma}-\tilde{\tilde{P}}_{\sigma})-[\Sigma_{1\sigma}^{r}+\Sigma_{0\sigma}^{r}(\tilde{P}_{\sigma}-\tilde{\tilde{P}}_{\sigma})+(\tilde{Q}_{\sigma}-\tilde{\tilde{Q}}_{\sigma})]G_{\sigma\sigma}^{r} \qquad (48)$$

, where

$$\tilde{P}_{\sigma}(E)=i\int\frac{d\varepsilon}{2\pi}\frac{1}{E+\varepsilon-\varepsilon_{\sigma}-\varepsilon_{\bar{\sigma}}-U+i\hbar/\tau_{\bar{\sigma}}}F_{\bar{\sigma}}(\varepsilon)\times$$
$$\times\left[G_{\bar{\sigma}\bar{\sigma}}^{r}(\varepsilon)\left(\Sigma_{0\bar{\sigma}}^{r}(\varepsilon)+\tilde{\Sigma}_{3\sigma}^{r}(E)\right)-G_{\bar{\sigma}\bar{\sigma}}^{a}(\varepsilon)\left(\Sigma_{0\bar{\sigma}}^{a}(\varepsilon)+\tilde{\Sigma}_{3\sigma}^{r}(E)\right)\right] \qquad (48\text{-}1)$$

$$\tilde{\tilde{P}}_{\sigma}(E)=i\int\frac{d\varepsilon}{2\pi}\frac{1}{E-\varepsilon-\varepsilon_{\sigma}+\varepsilon_{\bar{\sigma}}+i\hbar/\tau_{\bar{\sigma}}}F_{\bar{\sigma}}(\varepsilon)\times$$
$$\times\left[G_{\bar{\sigma}\bar{\sigma}}^{r}(\varepsilon)\left(\Sigma_{0\bar{\sigma}}^{r}(\varepsilon)-\tilde{\tilde{\Sigma}}_{3\sigma}^{r}(E)\right)-G_{\bar{\sigma}\bar{\sigma}}^{a}(\varepsilon)\left(\Sigma_{0\bar{\sigma}}^{a}(\varepsilon)-\tilde{\tilde{\Sigma}}_{3\sigma}^{r}(E)\right)\right] \qquad (48\text{-}2)$$

$$\tilde{Q}_{\sigma}(E)=i\int\frac{d\varepsilon}{2\pi}\frac{1}{E+\varepsilon-\varepsilon_{\sigma}-\varepsilon_{\bar{\sigma}}-U+i\hbar/\tau_{\bar{\sigma}}}F_{\bar{\sigma}}(\varepsilon)\times$$
$$\times\left[G_{\bar{\sigma}\bar{\sigma}}^{r}(\varepsilon)\Sigma_{0\bar{\sigma}}^{r}(\varepsilon)\left(\Sigma_{0\bar{\sigma}}^{r}(\varepsilon)+\tilde{\Sigma}_{3\sigma}^{r}(E)\right)-G_{\bar{\sigma}\bar{\sigma}}^{a}(\varepsilon)\Sigma_{0\bar{\sigma}}^{a}(\varepsilon)\left(\Sigma_{0\bar{\sigma}}^{a}(\varepsilon)+\tilde{\Sigma}_{3\sigma}^{r}(E)\right)\right] \qquad (48\text{-}3)$$

$$\tilde{\tilde{Q}}_{\sigma}(E)=i\int\frac{d\varepsilon}{2\pi}\frac{1}{E-\varepsilon-\varepsilon_{\sigma}+\varepsilon_{\bar{\sigma}}+i\hbar/\tau_{\bar{\sigma}}}F_{\bar{\sigma}}(\varepsilon)\times$$
$$\times\left[G_{\bar{\sigma}\bar{\sigma}}^{r}(\varepsilon)\Sigma_{0\bar{\sigma}}^{r}(\varepsilon)\left(\Sigma_{0\bar{\sigma}}^{r}(\varepsilon)-\tilde{\tilde{\Sigma}}_{3\sigma}^{r}(E)\right)-G_{\bar{\sigma}\bar{\sigma}}^{a}(\varepsilon)\Sigma_{0\bar{\sigma}}^{a}(\varepsilon)\left(\Sigma_{0\bar{\sigma}}^{a}(\varepsilon)-\tilde{\tilde{\Sigma}}_{3\sigma}^{r}(E)\right)\right] \qquad (48\text{-}4)$$

$$\tilde{\Sigma}_{i\sigma}^{r}(E)=\sum_{\beta}\int\frac{d\varepsilon}{2\pi}\Gamma_{\bar{\sigma}}^{\beta}A_{\beta}^{(i)}(\varepsilon)\frac{1}{E+\varepsilon-\varepsilon_{\sigma}-\varepsilon_{\bar{\sigma}}-U+i\hbar/\tau_{\bar{\sigma}}}$$
$$\tilde{\tilde{\Sigma}}_{i\sigma}^{r}(E)=\sum_{\beta}\int\frac{d\varepsilon}{2\pi}\Gamma_{\bar{\sigma}}^{\beta}A_{\beta}^{(i)}(\varepsilon)\frac{1}{E-\varepsilon-\varepsilon_{\sigma}+\varepsilon_{\bar{\sigma}}+i\hbar/\tau_{\bar{\sigma}}} \qquad (48\text{-}5)$$

$$\Sigma_{i\sigma}^{r}(E)=\tilde{\Sigma}_{i\sigma}^{r}(E)+\tilde{\tilde{\Sigma}}_{i\sigma}^{r}(E) \qquad (48\text{-}6)$$

Where $i=1,3$, $A_{\beta}^{(1)}(\varepsilon)=f_{\beta}(\varepsilon)$ and $A_{\beta}^{(3)}(\varepsilon)=1$. $F_{\sigma}(\varepsilon)$ denotes the pseudo-equilibrium distribution function and is defined as

$$F_{\sigma}(\varepsilon)=\frac{\Gamma_{\sigma}^{L}f_{L}(\varepsilon)+\Gamma_{\sigma}^{R}f_{R}(\varepsilon)}{\Gamma_{\sigma}^{L}+\Gamma_{\sigma}^{R}} \qquad (49)$$

It seems to be a weight-averaged function with half-width functions of the left lead and the right one. In Refs. 32,33, instead of such the 'pseudo-equilibrium' distribution function the equilibrium one appears, because they used the fluctuation-dissipation theorem that is valid for the equilibrium state in order to calculate the correlation terms $\langle A^\dagger B \rangle$.

Unlike Refs. 32,33, remembering that the correlation term $\langle A^\dagger B \rangle$ is related to the lesser Green function $\langle\langle B; A^\dagger \rangle\rangle^<_E$ with

$$\langle A^\dagger B \rangle = -i \int \frac{dE}{2\pi} \langle\langle B; A^\dagger \rangle\rangle^<_E \tag{50}$$

, we adopted the EOM technique for lesser Green functions, and $G^<_{\sigma\sigma}$ was derived as

$$G^<_{\sigma\sigma}(E) = -F_\sigma(E)[G^r_{\sigma\sigma}(E) - G^a_{\sigma\sigma}(E)] \tag{51}$$

by using Ng *ansatz*[30]. We regard that Eq. (51) would be the generalization of the fluctuation-dissipation theorem for the pseudo-equilibrium state.

Eq. (11) and (48) form the closed set of equations, and from those the QD retarded Green function $G^r_{\sigma\sigma}$ is given by

$$G^r_{\sigma\sigma} = \frac{E - \varepsilon_\sigma - \Sigma^r_{03\sigma} - U[1 - \langle n_{\bar\sigma} \rangle + (\tilde{P}_\sigma - \tilde{\tilde{P}}_\sigma)]}{(E - \varepsilon_\sigma - \Sigma^r_{0\sigma})(E - \varepsilon_\sigma - U - \Sigma^r_{03\sigma}) + U[\Sigma^r_{1\sigma} + \Sigma^r_{0\sigma}(\tilde{P}_\sigma - \tilde{\tilde{P}}_\sigma) + (\tilde{Q}_\sigma - \tilde{\tilde{Q}}_\sigma)]} \tag{52}$$

On the other hand, the DOS for electrons in QD is defined[22] by

$$DOS(E) = -\frac{1}{\pi} \mathrm{Im} G^r_{\sigma\sigma}(E) \tag{53}$$

and the current Eq. (5) is written as follows using Ng *ansatz*[30].

$$I_\sigma = \frac{ie}{\hbar} \int \frac{d\varepsilon}{2\pi} \frac{\Gamma^L_\sigma \Gamma^R_\sigma}{\Gamma^L_\sigma + \Gamma^R_\sigma} (G^r_{\sigma\sigma} - G^a_{\sigma\sigma})[f_L - f_R] \tag{54}$$

In order to calculate the current Eq. (54) we must find out $G^r_{\sigma\sigma}$ and to do this we must find out the average occupation number of electrons $\langle n_{\bar\sigma} \rangle$ This is found by the definition of the lesser Green function $G^<_{\sigma\sigma}$.

$$\langle n_\sigma \rangle = \int \frac{d\varepsilon}{2\pi} \mathrm{Im} G^<_{\sigma\sigma} = i \int \frac{d\varepsilon}{2\pi} \frac{\Gamma^L_\sigma f_L + \Gamma^R_\sigma f_R}{\Gamma^L_\sigma + \Gamma^R_\sigma} (G^r_{\sigma\sigma} - G^a_{\sigma\sigma})$$

$$= i \int \frac{d\varepsilon}{2\pi} F_\sigma(\varepsilon) (G^r_{\sigma\sigma}(\varepsilon) - G^a_{\sigma\sigma}(\varepsilon)) \tag{55}$$

Therefore, to find $G^r_{\sigma\sigma}$, Eqs. (48-1~6),(51),(52) and (55) must be calculated self-consistently.

## IV. Numerical Results

For simplicity, the electron band in the leads is assumed to be independent of the energy and extends from $-D$ below Fermi level (bottom band edge) up to $D$ above Fermi level (top band edge). The energy integrals will cut off at $-D \leq E \leq D$, i.e., will be limited to the electron band. For positive (negative) bias we assume the electrostatic potential of the left(right) electrode equal to zero. In other words, the electrochemical potential of the drain electrode is assumed to be zero, while that of the source electrode is shifted up by $|eV|$.

## A. Density of State

The main feature of the Kondo effect in QD is an appearance of the Kondo peak in the DOS near Fermi level. Fig. 1 shows the height of the Kondo peak (value of DOS at Fermi level) vs. temperature $T/T_K$.

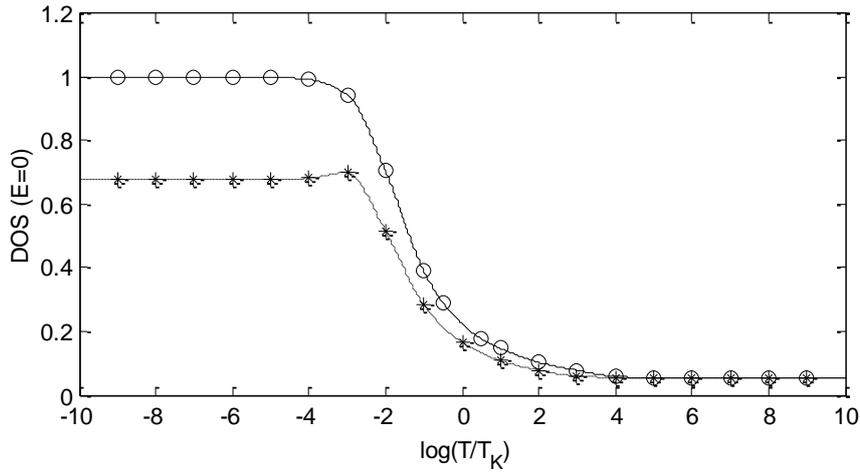

**Fig. 1. The height of the Kondo peak vs. temperature**.

The solid and dashed lines are the result using the approximation developed by Meir *et al*[15]. and Lacroix[29] respectively. The parameters assumed for numerical calculations are: $\varepsilon_0 = -0.6$, $U = 4$, $\Gamma_L = \Gamma_R = 0.1$, $D = 2$.

As shown in Fig. 1, the approximation worked out by Meir *et al*[15] is valid only for the high-temperature regime $T \gg T_K$, but gives rise to the distorted result in the low-temperature regime $T \ll T_K$.

When the temperature decreases, there appears the Kondo peak in DOS near Fermi level. The height of the peak increases with decreasing temperature. Furthermore, the height of the Kondo peak in the very low temperature $T \ll T_K$ approaches to the maximum 1. This value reaches exactly to 1 at $T = 0$. The result agrees with (numerical renormalized group)NRG[38] and (non-crossing approximation)NCA[39] results.

## B. Zero-Bias Conductance

Fig. 2 shows the zero-bias conductance vs. $T \ll T_K$. The conductance follows the universal behavior in a wide range of temperature. Deviation from this universality is observed only for $T \gg T_K$, as expected. We point out at very low- temperature $T/T_K \ll 1$ the universal curve has the expected Fermi- liquid behavior, i.e. $(G - G_0)/G_0 \propto T^2$, while at higher temperature $T/T_K \approx 1$ the conductance is proportional to $\ln(T/T_K)$. The dependence curve of the conductance on temperature is in agreement with results of NRG[40], NCA[34,35] and non-equilibrium renormalized theory(NRPT)[36].

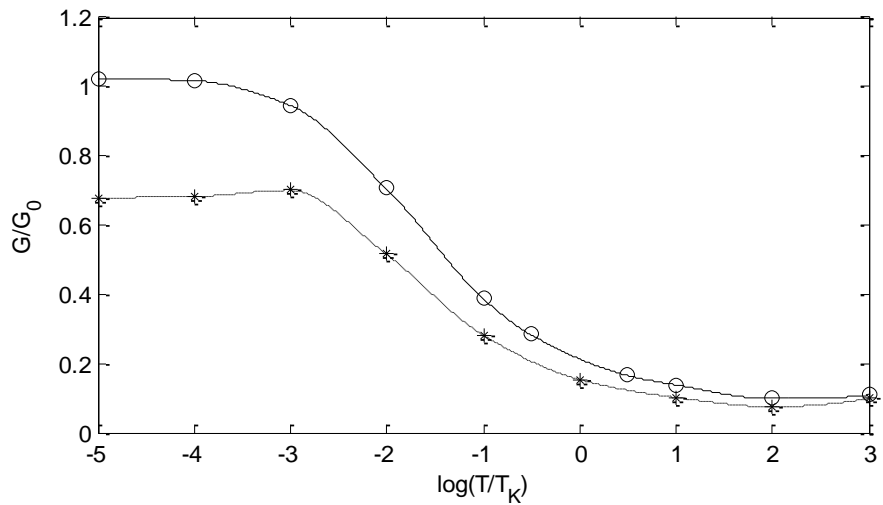

**Fig. 2. The zero-bias conductance vs. temperature.**

The parameters, the solid and dashed lines are same as in Fig. 1.

### V. Conclusion

In this paper, we have studied theoretically the non-equilibrium Kondo effect in the quantum dot within the whole range of temperature including the Kondo temperature by using the EOM method based on the non-equilibrium Green function technique. We have taken the finitness of Coulomb correlation and the non-equilibrium effect into account by calculating the correlation terms emerged from the decoupling approximation using the EOM method for the lesser Green function. The results would be the generalization into the pseudo-equilibrium state of Refs. 32,33 and can be used to describe a non-equilibrium state under the bias voltage which is not so large. The numerical results agree with the results of NCA, NRG and NRPT, etc.